\documentclass[final,3p,times,twocolumn,authoryear]{elsarticle}

\usepackage{lineno,hyperref}
\usepackage{graphicx}
\usepackage{CJKutf8}
\usepackage{txfonts}
\usepackage{xcolor}
\usepackage{lipsum}
\usepackage{tabularx}
\usepackage{siunitx}
\usepackage{amsmath}
\usepackage{footnote}
\usepackage{soul}
\usepackage{multirow}
\usepackage{multicol}
\newcommand{\insight}{\textit{Insight}-HXMT}
\newcommand{\gbm}{\textit{Fermi}/GBM}
\newcommand{\bat}{\textit{Swift}/BAT}

\newcommand{\cmmnt}[1]{}
\DeclareSIUnit\gauss{G}
\DeclareSIUnit\year{yr}
\DeclareSIUnit\erg{erg}
\newcommand{\ws}{\omega_{\mathrm{s}}}

\modulolinenumbers[5]

\journal{Journal of \LaTeX\ Templates}

\biboptions{authoryear}

\begin{document}

\begin{frontmatter}

\title{\insight{} insight into switch of the accretion mode: the case of the X-ray pulsar 4U 1901+03}

\author[a,b]{Y. L. Tuo\corref{cor1}}
\ead{tuoyl@ihep.ac.cn}
\author[c]{L. Ji}
\author[d,e]{S. S. Tsygankov}
\author[f]{T. Mihara}

\author[a]{L. M. Song\corref{cor1}}
\ead{songlm@ihep.ac.cn}
\author[a]{M. Y. Ge}
\author[d]{A. Nabizadeh}
\author[a]{L. Tao}
\author[a]{J. L. Qu}
\author[a,b]{Y. Zhang}
% ...

\address[a]{Key Laboratory of Particle Astrophysics, Institute of High Energy Physics, Chinese Academy of Sciences, Beijing 100049, China}
\address[b]{University of Chinese Academy of Sciences, Chinese Academy of Sciences, Beijing 100049, China}
\address[c]{Institut f\"ur Astronomie und Astrophysik, Kepler Center for Astro and Particle Physics, Eberhard Karls Universit\"at, 72076 T\"ubingen, Germany}
\address[d]{Department of Physics and Astronomy, FI-20014 University of Turku, Turku, Finland}
\address[e]{Space Research Institute of the Russian Academy of Sciences, Profsoyuznaya Str. 84/32, Moscow 117997, Russia}
\address[f]{High Energy Astrophysics Laboratory, Institute of Physical and Chemical research RIKEN, Wako, Saitama 351-0198, Japan}
\address[g]{Department of Astronomy, Beijing Normal University, Beijing 100088, China}
\address[h]{Department of Astronomy, Tsinghua University, Beijing 100084, China}
\address[i]{Department of Physics, Tsinghua University, Beijing 100084, China}
\address[j]{Department of Engineering Physics, Tsinghua University, Beijing 100084, China}
\address[k]{Key Laboratory of Space Astronomy and Technology, National Astronomical Observatories, Chinese Academy of Sciences, Beijing 100012, China}

\begin{abstract}
We use the \insight{} data collected during the 2019 outburst from X-ray pulsar 4U 1901+03 to complement the orbital parameters reported by \gbm{}. Using the \insight{}, we examine the correlation between the derivative of the intrinsic spin frequency and bolometric flux based on accretion torque models. 
It was found that the pulse profiles significantly evolve during the outburst. The existence of two types of the profile's pattern discovered in the \insight{} data indicates that this source experienced transition between a super-critical and a sub-critical accretion regimes during its 2019 outburst. Based on the evolution of the pulse profiles and the torque model, we derive the distance to 4U 1901+03 as $12.4\pm0.2\,\mathrm{kpc}$.
\end{abstract}


\end{frontmatter}

\nolinenumbers

\section{Introduction}\label{sec:intro}
Binary systems hosting neutron stars (NSs) are among the most powerful X-ray sources in our Galaxy. The binary systems with a massive stellar companion, usually an O or B star, are characterized as high mass X-ray binaries (HMXBs). The X-ray emission from HMXBs is due to the transfer of matter from the companion star and the accretion onto the NS. If NS in such system possesses strong magnetic field (of the order of $10^{12}$~G or even stronger), they exhibit themselves as X-ray pulsars (XRP; see \cite{walter2015high} for a recent review).
The properties of emission registered during bright outbursts from the transient XRPs provide an insight into the physics of accretion and transfer of torque to the NS \citep{ghosh1979accretion,wang1987disc,wang1996location,zhang2019insight}.\par
The high-mass X-ray binary 4U 1901+03 was discovered by \textit{Uhuru} and \textit{Vela 5B} in 1970-1971 \citep{forman1976uhuru,priedhorsky1984long}.
\cite{galloway2005discovery} reported results of the \textit{RXTE} observations performed during an outburst in 2003 when the peak flux reached  $8\times10^{-9}\,\mathrm{erg}\,\mathrm{cm}^{-2}\,\mathrm{s}^{-1}$ in 2.5--25\,keV.
They detected a clear pulsations with a spin period of $\sim$ 2.763\,s, and obtained the orbital ephemeris, where orbital period is 22.58 days, based on the Doppler effect caused by the motion in the binary system.
After a long quiescence state, \textit{MAXI}/GSC found a new outburst in 2019, with a peak flux of $\sim$ 200\,mCrab \citep{nakajima2019maxi}.
\par
In this paper we report the results of the timing analysis of emission from 4U 1901+03, such as binary parameters, making use of the \insight{} data as a complement to \gbm{} results (P. Jenke, in prep). The pulse profiles of the NS at different luminosity states are reported as well. In addition, we investigate the spin-up rate and the accreting torque during the outburst, and estimate the distance of the source according to theoretical torque models. The methods for the data reduction and analysis are presented in section \ref{sec:analysis}. The obtained results are discussed in the frame of theoretical models in section \ref{sec:discussion}.

\section{Data analysis and results}\label{sec:analysis}
To analyze the evolution of intrinsic spin frequency and flux during the recent outburst, we utilize the data from the {\it Fermi}/Gamma-ray Burst Monitor (GBM), the {\it Swift}/Burst Alert Telescope (BAT), and the {\it Insight}-Hard X-ray Modulation Telescope (\insight{}). 

\par
\bat{} is a hard X-ray transient monitor providing near real-time coverage of the X-ray sky in the energy range 15--50\,keV \citep{krimm2013swift}. It provides the flux evolution of 4U 1901+03 in 15--50\,keV during the whole outburst. For the \bat{} data\footnote{https://swift.gsfc.nasa.gov/results/transients/index.html} we selected the Daily light curves from MJD 58520 to MJD 58649. Also we performed multiple follow-up observations of the source with \insight{} started from MJD 58573 to MJD 58644.8. With 36 observations carried out by the \insight{}, the total exposure time is about 150\,ks.
The \insight{} provides a broad band energy coverage in 1--250\,keV \citep{zhang2014introduction,zhangoverview}. The time resolution of three instruments, the high energy instrument (HE) \citep{liu2019high}, the medium energy instrument (ME) \citep{cao2019medium}, and the low energy instrument (LE) \citep{chen2019low} on-board the \insight{} are \SI{2}{\us}, \SI{20}{\us}, and \SI{1}{\ms}, respectively. The \insight{} provides information about the flux and the temporal properties of 4U 1901+03. The frequency evolution and the frequency derivative could have continuous results in the case that the \gbm{} provides a continuous observation.

\subsection{Data reduction}
Scientific data for the timing and spectral results, are obtained from the reduction of the \insight{} raw data.
The methods of data reduction for the \insight{} were introduced in previous publications  \citep[see e.g.,][]{Huang2018,Chen2018}. We summarize the procedures of using the \insight{} Data Analysis Software package (\texttt{HXMTDAS}) version 2.01 here:
\begin{enumerate}
	\item Use the commands \texttt{hepical}, \texttt{mepical}, \texttt{lepical} in \texttt{HXMTDAS} to calibrate the photon events from the raw data according the the Calibration Database (CALDB) of the \insight{}.
	\item Select the good time intervals (GTIs) for calibrated photons, using \texttt{hegtigen}, \texttt{megtigen}, and \texttt{legtigen}.
	\item Extract the good events based on the GTIs using the commands \texttt{hescreen}, \texttt{mescreen}, and \texttt{lescreen}.
	\item Generate spectra for selected photons using the commands \texttt{hespecgen}, \texttt{mespecgen}, and \texttt{lespecgen}.
	\item Generate the background spectra based on the emission detected by blind detectors using the commands \texttt{hebkgmap}, \texttt{mebkgmap}, and \texttt{lebkgmap}.
	\item Generate the response matrix files required for spectral analysis using the commands \texttt{herspgen}, \texttt{merspgen}, and \texttt{lerspgen}.
\end{enumerate}
We apply strict criteria to create the GTI file used for spectral analysis. We set the parameters Earth elevation angle (ELV) greater than 10 degrees, the cutoff rigidity (COR) greater than 8\,GeV, the offset angle from the pointing source (ANG\_DIST) less than 0.04 degrees. We also exclude the photons collected  300\,s before enter and after exit the South Atlantic Anomaly (SAA) region. However, in order to optimize the count statistics required for timing analysis, less strict criteria for filtering events are applied. Particularly, we select the time intervals when $\mathrm{ELV}>0$ and the satellite is not in the SAA region. The total effective exposure time for \insight{} after data screening is about 84.3\,ks.
\par
The arrival times of photons from the \insight{} data are corrected to the Solar system barycenter using the \texttt{HXMTDAS} commands \texttt{hxbary}. The coordinates of the source are taken as (J2000): RA=$19\mathrm{h}\, 03\mathrm{m}\,39.42\mathrm{s}$, Dec=$+03\mathrm{d}\,12'\, 15.8''$ \citep{halpern2019chandra}.
\par

\subsection{Timing analysis}
\label{subsec:timing}

\begin{figure}
    \centering\
    \includegraphics[width=0.5\textwidth]{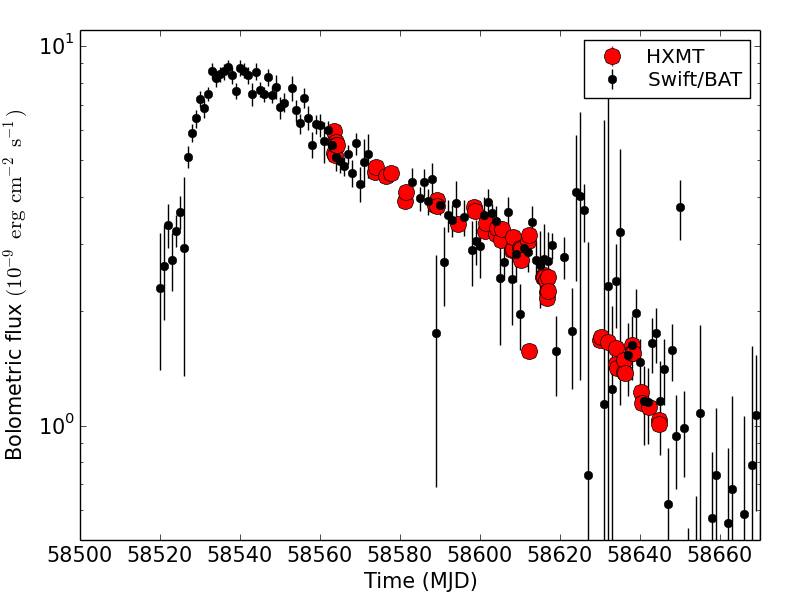}
    \caption{The bolometric lightcurve of 4U 1901+03 as seen by \textit{Swift}/BAT (black points), obtained by scaling the count rate to match the \insight{} flux in 0.1--150\,keV band (red circles). The \insight{} flux was obtained through the spectral fitting. }
    \label{fig:f_bol}
\end{figure}

In Figure \ref{fig:f_bol} the bolometric lightcurve of 4U 1901+03 based on the \bat{} data is presented. The count rate provided by \bat{} in 15--50 keV band was converted to match the flux calculated using spectra from \insight{}. Due to dependence of spectral shape on the source flux, one conversion factor is not enough to make the \bat{} flux matches the \insight{} fluxes well. Thus the conversion factors were estimated for different time intervals as $1.9\times 10^{-7}\,\mathrm{erg}\,\mathrm{cnt}^{-1}$ for the data before MJD 58580, $1.7\times 10^{-7}\,\mathrm{erg}\,\mathrm{cnt}^{-1}$ for the data between MJD 58580 and MJD 58597, $1.8\times 10^{-7}\,\mathrm{erg}\,\mathrm{cnt}^{-1}$ for the data after MJD 58597. These factors convert the \bat{} count rate in unit of $\mathrm{cnt}\,\mathrm{cm}^{-2}\,\mathrm{s}^{-1}$ to the flux in the units of $\mathrm{erg}\,\mathrm{cm}^{-2}\,\mathrm{s}^{-1}$. 
The background subtracted spectra from the \insight{} data were analyzed in the 1--150\,keV band.
The more detailed spectral analysis based on the \insight{} data is ongoing and will be published elsewhere. 

The bolometric flux was estimated from fitting the broadband \insight{} spectrum in 0.1--150\,keV band as follows:
\begin{enumerate}
    \item Fit each spectrum in 1--150\,keV with a model \texttt{TBabs*cutoffpl} in \texttt{XSPEC (v12.10.0c)}.
    \item Expand the energy range to 0.1--150\,keV for response matrix using the command \texttt{energies}. 
    \item Freeze the best fitted normalization value of the model cutoff power law.
    \item Add \texttt{cflux} component in order to calculate an unabsorbed flux from the \texttt{cutoffpl} model in 0.1--150\,keV energy band. 
\end{enumerate}
The photon index and the e-folding energy of exponential rolloff in the \texttt{cutoffpl} for the very first observation of \insight{} are 0.46 and 7.08 respectively. The flux in 0.1--150\,keV can serve as a good estimate for the bolometric flux of an XRP. 

\begin{figure}
    \centering
    \includegraphics[width=0.5\textwidth]{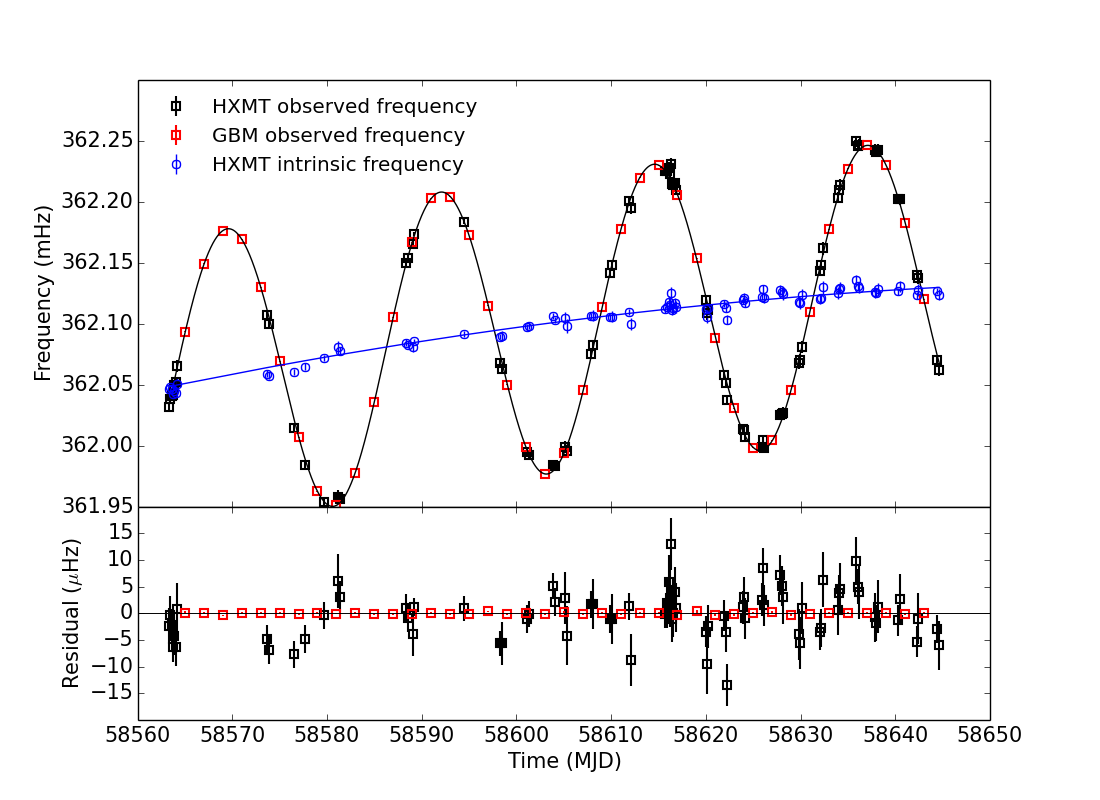}
    \caption{The observed frequencies of 4U 1901+03 obtained from the \gbm{} and \insight{} data are shown with red and black squares in the upper panel, respectively. The intrinsic spin frequencies for the \insight{} data set are presented with blue circles. The blue line indicates the best fit polynomial function to the intrinsic frequency. The residuals between frequency model and data are presented in the bottom panel.
    }\label{fig:f0}
\end{figure}

The preliminary orbital parameters of 4U 1901+03 are presented in the GBM Accreting Pulsar  Histories project\footnote{https://gammaray.msfc.nasa.gov/gbm/science/pulsars.html}. Here we report the spin frequency obtained by the \insight{} to complement the available \gbm{} measurements (P. Jenke, in prep). The spin period in each \insight{} observation was calculated using an epoch-folding technique \citep{leahy1987searches}. Uncertainty for the spin period were roughly estimated from the width of $\chi^2$ distribution for the trial periods.
The observed frequencies obtained by the \insight{} combine the intrinsic spin frequency of the NS and effect of the Doppler shift due to the binary motion (see black squares in Figure \ref{fig:f0}). As seen from Figure \ref{fig:f0} the frequency evolution is modulated by an almost sinusoidal function. To calculate the intrinsic spin frequency of the NS we apply method described in  \cite{galloway2005discovery}. Here we outline the method of the corresponding calculations. The observed frequencies could be written as,
\begin{equation}
\label{eq:obs_fre}
\begin{aligned}
f(t) = & f_{\mathrm{spin}}(t)  \\ 
& - \frac{2\pi f_0 a_{\mathrm{X}}\sin{i}}{P_{\mathrm{orb}}}(\cos{l}+ g\sin{2l} + h\cos{2l}),
\end{aligned}
\end{equation}
where the first term $f_\mathrm{spin}(t)$ is the intrinsic spin frequency of the pulsar, and the second term is the frequency modulation due to the binary motion. $f_0$ is a constant approximating $f_\mathrm{spin}(t)$, $a_\mathrm{X}\sin{i}$ is the projected orbital semi-major axis in units of light seconds, $i$ is the system inclination, $P_\mathrm{orb}$ is the orbital period in unit of days. $g=e\sin{\omega}$, $h=e \cos{\omega}$ are functions of eccentricity $e$ and longitude of periastron $\omega$. And $l=2\pi(t-T_{\pi/2})/P_{\mathrm{orb}}+\pi/2$ is the mean longitude. The reference time $T_\mathrm{\pi/2}$ is when the mean longitude $l=\pi/2$, where the NS is behind the companion. 

The $f_{\mathrm{spin}}(t)$ was described by a third order polynomial function,

\begin{equation}
    f_{\mathrm{spin}}(t) = f_0 + \dot{f}(t-t_0) + \frac{1}{2}\ddot{f}(t-t_0)^2 + \frac{1}{6}\dddot{f}(t-t_0)^3 
\end{equation}
where $f_0$ is the frequency at reference time $t_0$. We arbitrary select $t_0$ as the beginning of the \insight{} observation of the source. And $\dot{f}$, $\ddot{f}$, and $\dddot{f}$ are the first, second, and third order derivatives of intrinsic frequency, respectively.
Using Equation \ref{eq:obs_fre} we fit the joint data set consisting of the \gbm{} and \insight{} data. 
In Figure \ref{fig:f0}, the observed frequency from the \insight{} and \gbm{} data are plotted in black and red squares, respectively. The blue circles in the top panel correspond to the intrinsic frequency from the \insight{} data. The bottom panel shows the residuals between $f(t)$ model and observed frequencies. The best fit results are listed in Table \ref{tab:orbpar}, where the values in parentheses are the uncertainties. The reduced $\chi^2$ is 1.41 with 109 d.o.f. As can be seen the errors on frequency in the \gbm{} data are about $10^{-7}$\,Hz, while the \insight{} results have errors about one magnitude greater. This is due to quite short exposure time of each \insight{} observation of a few thousands seconds. At the same time the \gbm{} provides an almost uninterrupted coverage of the outburst. We consider the weights for frequency data based on their error when fitting the orbital parameters.

\par

\begin{table}[t]
\caption{Temporal parameters for 4U 1901+03}\label{tab:orbpar}
\centering
\begin{tabular}{lcc }
\hline
\hline
    Parameter & GBM results  & GBM+HXMT results \\
    \hline
    $P_\mathrm{orb}$ (days)             & 22.534777  & 22.534571(11) \\
    $a_{\mathrm{X}}\sin{i}$ (lt\,s)      & 104.343  & 104.236(39)   \\
    $e$                          & 0.0150          &  0.01443(35) \\
    $\omega$ (deg)                     & 220.10    &  220.034(25)  \\
     $T_{\pi/2}$ (MJD)                  & \multicolumn{2}{c}{55927.26871}\\
     $f_0$ (Hz)  & -                                  &  0.36204843(13)   \\
     $\dot{f}$ (\si{\hertz\,s^{-1}}) & -                  &  1.833(16)e-11\\
     $\ddot{f}$ (\si{\hertz\,s^{-2}})& -                 & -1.89(11)e-18  \\
     $\dddot{f}$ (\si{\hertz\,s^{-3}})& -                 & -1(3)e-26  \\
     $t_0$ (MJD) & -                             &  58563.3290883878 \\
     $\chi^2$/dof    & - & 154/109 \\
  \hline
\end{tabular}
\end{table}

\begin{figure}[t]
    \centering
    \includegraphics[width=0.5\textwidth]{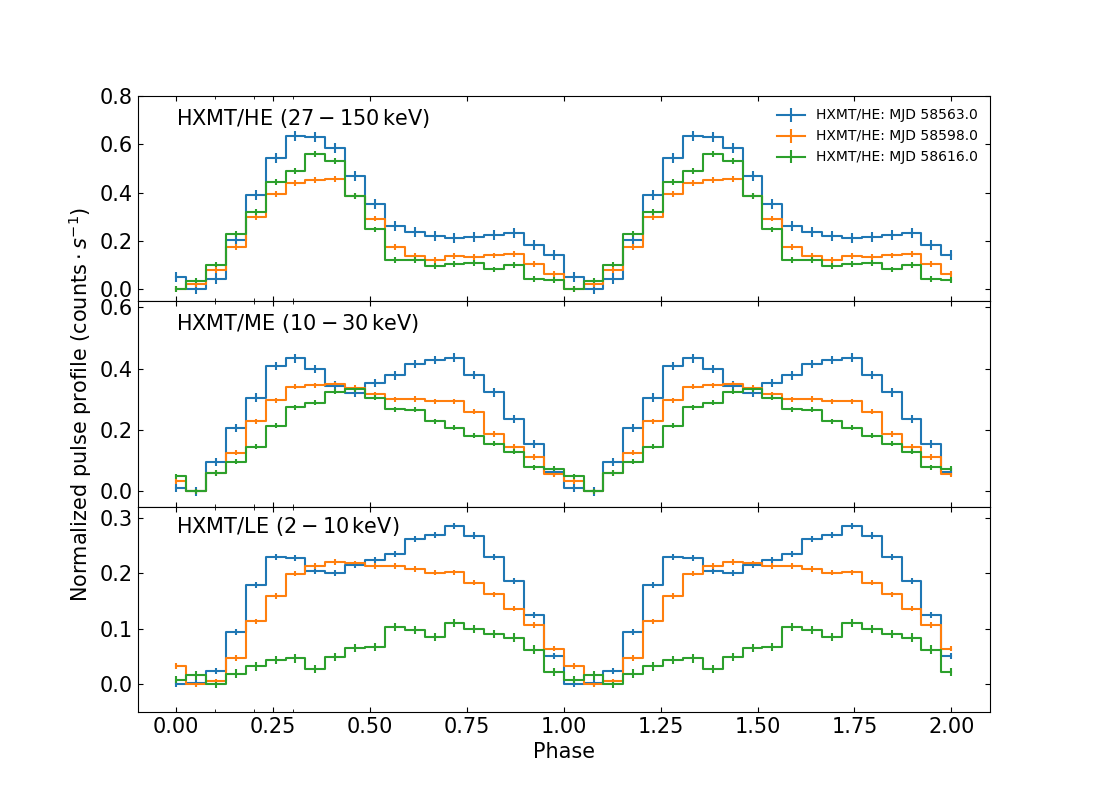}
    \caption{The evolution of pulse profiles of 4U 1901+03 in three different energy bands with time as observed by \insight{}. The whole set of observations of the \insight{} is split into three time intervals. The blue line are the pulse profiles in MJD 58563 to MJD 58598. The green line are the pulse profiles in MJD 58598 to MJD 58616. The pulse profiles observed in the last time interval MJD 58616 to MJD 58644 are presented with red lines. The profiles were normalized by subtracting the lowest count rate in profiles. The pulse profiles in three energy band, 27--150\,keV, 10--30\,keV, and 2--10\,keV, are plotted in upper, middle, and the bottom panels, respectively.}
    \label{fig:profile_insight}
\end{figure}

The obtained spin frequencies of the NS were utilized to produce pulse profiles at different luminosities. For plotting purposes the profiles in different observations were aligned together using cross-correlation function. We generate the profiles for each observation, the pulse profiles varies in time. To study the evolution of profiles in time and accumulate profiles for higher significance, we separate the \insight{} data into three time intervals, MJD 58563-58598, MJD 58598-58616, MJD 58616-58649. For clarity, we name these three intervals as $\mathrm{T}_1$, $\mathrm{T}_2$, and $\mathrm{T}_3$ hereafter. The resulting profiles in three time intervals obtained using the data from all three instruments of the \insight{} are shown in Figure \ref{fig:profile_insight}. 
\par
As can be seen, the shape of the pulse profiles evolves with energy. In hard energy band 27--150\,keV covered by the HE instrument the profile remains narrow single peaked, while for the ME and LE instrument, the pulse profiles vary from double peak shapes to broad single peaked. The pulse profile depends on the flux as well. The averaged fluxes in time intervals T1, T2, and T3 are $4.7\times10^{-9}\,\mathrm{erg}\,\mathrm{cm}^{-2}\,\mathrm{s}^{-1}$, $3.0\times10^{-9}\,\mathrm{erg}\,\mathrm{cm}^{-2}\,\mathrm{s}^{-1}$, and $1.7\times10^{-9}\,\mathrm{erg}\,\mathrm{cm}^{-2}\,\mathrm{s}^{-1}$, respectively.
In the $\mathrm{T}_1$ interval, pulse profile in 10--30\,keV range is double peaked. When the luminosity decreases, the pulse profile change to a broad single peaked in $\mathrm{T}_2$ and $\mathrm{T}_3$ intervals. For the profiles in 2--10\,keV the variations are similar. Particularly, in the $\mathrm{T}_1$ interval a double peaked shape is detected. And single peaked profiles are detected in the $\mathrm{T}_2$ and $\mathrm{T}_3$ intervals.
\par
The pulsed fraction of each profile are calculated by $\Sigma$(flux per bin - minimum per bin) divided by the total flux. In the time interval T1, the pulsed fractions for LE, ME, and HE are 0.12, 0.032, 0.015, respectively. Similarly, in the time interval T2, the pulsed fractions are 0.068, 0.032, 0.017. In the time interval T3, the pulsed fractions are 0.008, 0.02, 0.01. The pulsed fractions decrease with energy in the time interval T1, and T2. In the time interval T3, the pulsed fraction reaches peak in the energy band of ME.

\section{Discussion and conclusion}
\label{sec:discussion}

In this work, we investigate the temporal evolution of the coherent X-ray pulsations shown by 4U 1901+03 during its outburst in 2019, using the \gbm{} frequency results and the whole data set collected by the \insight{}. The orbital parameters and the intrinsic timing parameters are obtained (see Table.\ref{tab:orbpar}). The pulse profile evolution with luminosity and energy observed by the \insight{} are presented as well. Using the long-term monitoring of the source we searched for a possible propeller effect reported by \cite{reig2016accretion} 150 days after the outburst in 2003. However, we didn't discover any sharp drops of the flux similar to another XRPs \citep{tsygankov2016propeller}. We also investigate the torque behavior during the outburst making use of the model in \cite{ghosh1979accretion} (GL model hereafter) and examine the correlation between the frequency derivative and the luminosity. 
\par

\subsection{Accretion torque}
The orbital parameters updated using the \gbm{} and \insight{} data provide us an opportunity to analyze the spin-up behavior of 4U 1901+03 excluding the Doppler effect from orbital modulation.
The spin evolution of the NS is driven by accretion torque during the outburst that can be written as:
\begin{equation}
    \label{eq:torque}
    \dot{f} = \frac{df}{dt} = \frac{N}{2\pi I},
\end{equation}
where $I$ is the effective moment of inertia of the NS, and $N$ is the total torque. GL model assumes a magnetically-threaded disk having effects on the NS. Since the accreting materials have a torque onto the NS, the frequency derivative of the NS and the X-ray luminosity follows the correlation (GL model):

\begin{equation}
\label{eq:fdot-L}
    \dot{f} = 5.0\times10^{-5} \mu_{30}^{2/7} n(\ws) R_6^{6/7} I_{45}^{-1} (\frac{M_{\mathrm{NS}}}{M_{\odot}})^{-3/7} L_{37}^{6/7} \, \si{\hertz\per\year},
\end{equation}
where $\mu_{30}$ is the neutron star magnetic dipole moment in the disk plane ($\mu = \frac{1}{2}BR^3$) in units of \SI{e30}{\gauss\cubic\centi\metre}, $B$ is the magnetic field at the pole, $R_6$ is the radius of the NS in units of \SI{e6}{\centi\metre}, $I_{45}$ is the moment of inertia of the NS in units of \SI{e45}{\gram\square\centi\metre}, $M_\mathrm{NS}$ is the mass of the NS in units of grams, the $M_{\odot}$ is the solar mass in units of grams, and the $L_{37}$ is the luminosity in units of \SI{e37}{\erg\per\second}. The dimensionless torque $n(\ws)$ could be estimated as (GL model):
\begin{equation}
\label{eq:nws}
    n(\ws) \approx 1.39 \{1-\ws [4.03(1-\ws)^{0.173} - 0.878]\}(1-\ws)^{-1} ,
\end{equation}
where $\ws$ is the fastness parameter \citep{elsner1977accretion}, 
\begin{equation}
\label{eq:ws}
\ws \equiv \Omega_{\mathrm{s}} / \Omega_{\mathrm{K}}(R_\mathrm{m})
\end{equation}
which is the ratio between the angular velocity of the NS and the Keplerian angular velocity at the magnetospheric radius.

\par
As matter accreting on the NS, the angular momentum is carried from the Keplerian accretion disk to the NS. In the case of XRPs, the inner radius of the disk is the magnetospheric radius ($R_\mathrm{m}$), where the accreting matter transfers the orbiting angular moment to the NS. Thus the angular momentum transported to the NS is $\dot{M}\sqrt{GMR_\mathrm{m}}$. Multiple theories estimated the relation of $R_\mathrm{m}$ to the Alf{\'v}en radius ($R_\mathrm{A}$), where ram pressure of the spherical freely in-falling matter equals the magnetic pressure \citep{davidson1973neutron,waters1989relation}:
\begin{equation}
\label{eq:rm}
    R_\mathrm{m}=\xi R_\mathrm{A}=\xi(2GM)^{-1/7}\mu^{4/7}\dot{M}^{-2/7} ,
\end{equation} 
where $\xi$ is constant between 0 and 1 \citep{ghosh1979accretion,wang1987disc,wang1996location}. We use the $\xi$ value from \cite{ghosh1979accretion}, where $\xi=0.52$. 
Substituting Equation \ref{eq:rm} into Equation \ref{eq:ws} one obtains:
\begin{equation}
    \label{eq:ws1}
\ws = 1.19 P^{-1} \dot{M}_{17}^{-3/7} \mu_{30}^{6/7} (M_{\mathrm{NS}}/M_{\odot})^{-5/7},
\end{equation}
where $P$ is the NS spin period in seconds and $\dot{M}_{17}$ is mass accretion rate in units of $10^{17}$~g~s$^{-1}$.
The dimensionless torque $n(\ws)\approx1.4$ during the 
\insight{} observation, indicating the NS in 4U 1901+03 is a slow rotator. 

\par

\begin{figure}
    \centering
    \includegraphics[width=0.48\textwidth]{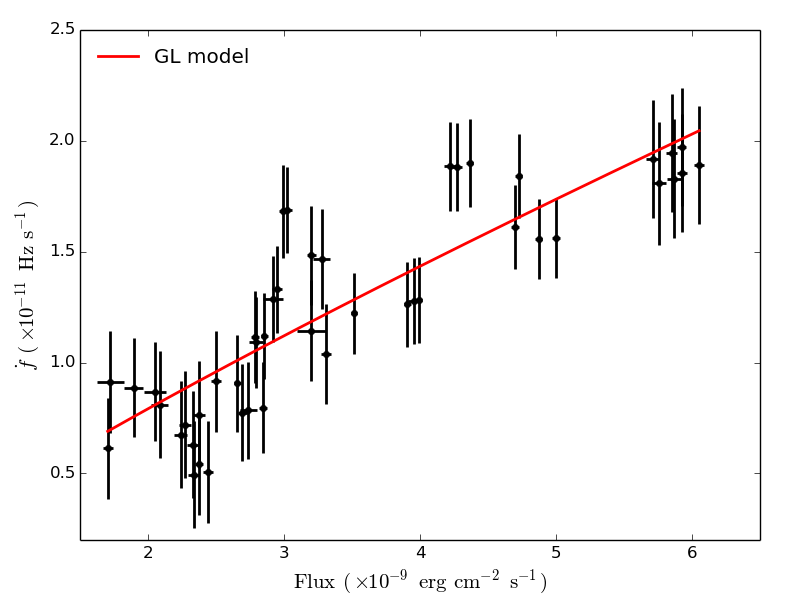}
    \caption{The correlation between frequency derivatives and bolometric flux as seen by \insight{}. The solid line is approximation with the GL model. }
    \label{fig:nudot-flux}
\end{figure}

To analyze the correlation between frequency derivatives ($\dot{f}$) and bolometric flux ($F$) using the GL model,
we calculate the representative $\dot{f}$ using the \gbm{} data covering time intervals with accurate flux measurements provided by \insight{}. The frequency derivatives $\dot{f}$ were obtained as $\Delta f/\Delta t$ in each consequent time intervals \citep{doroshenko2018orbit} using \gbm{} data corrected for the effects of the orbital motion. We select the midpoint of each time interval as representative time of the corresponding measurement. Since this time doesn't necessarily coincides with the flux measurement time, we have to interpolate the $\dot{f}(t)$ using linear interpolation method to match the representative time of both measurements. The errors on $\dot{f}$ are derived as the propagating errors of calculating $\Delta f/\Delta t$ and linear interpolation. The results are shown in Figure \ref{fig:nudot-flux}. The fitting procedure reveals the correlation $D = 15.935 \times B_{12}^{-1/6}\,[\mathrm{kpc}]$ according to the Equation \ref{eq:fdot-L}.

\begin{figure}
    \centering
    \includegraphics[width=0.5\textwidth]{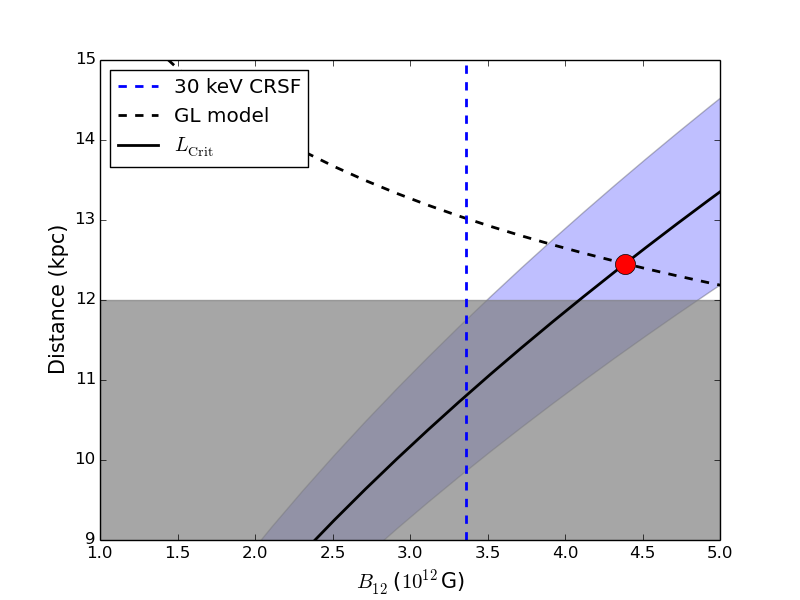}
    \caption{The distance -- magnetic field diagram for 4U 1901+03. The solid line is the correlation following from the critical luminosity. The optical observation constrain the distance greater than 12\,kpc (above the grey area). The red circle is the intersection with the torque model (shown with black dashed line), suggesting the distance of $12.4 \pm 0.2\,\mathrm{kpc}$. Vertical blue dashed line corresponds to the magnetic field derived from the cyclotron feature at 30 keV.}
    \label{fig:distance}
\end{figure}

The GL model links the observed frequency derivative with the magnetic field strength and mass accretion rate onto the NS. According to Equation \ref{eq:fdot-L}, the distance and the magnetic field correlation is plotted in black dashed line in Figure \ref{fig:distance}.
The possible cyclotron resonance scattering features (CRSFs) are located at 10\,keV \citep{reig2016accretion} and 30\,keV \citep{coley2019possible}. The fundamental cyclotron line energy is related to the magnetic field strength of the neutron star as $E_{\mathrm{cyc}}=11.6(1+z)^{-1}B_{12}$~keV,
where z is the surface gravitational red-shift, and $B_{12}$ is the magnetic field in the unit of $10^{12}$ G \citep[e.g.,][]{staubert2019cyclotron}. We assume $z=0.3$ for a typical neutron star mass of 1.4\,$M_{\odot}$ and $R=10^6\,\mathrm{cm}$.
The magnetic field suggested by two different CRSFs are $1.1\times10^{12}\,\mathrm{G}$ and $3.4\times10^{12}\,\mathrm{G}$, respectively. Using these magnetic field strengths the distances to 4U 1901+03 according to the GL model can be estimated as 15.6\,kpc and 13.0\,kpc. They are consistence with an expected large distance greater than 12\,kpc according to the optical observations \citep{strader2019optical}. To verify which distance is more reliable, we take the evolution of pulse profiles during the outburst into account. 

\subsection{Critical luminosity}
The evolution of pulse profiles during the outburst can be interpreted due to a transition of the source through the critical luminosity \citep{chen2008study, becker2012spectral,mushtukov2015critical,weng2019nustar,10.1093/mnras/stz2745, doroshenko2019hot}. In the bright state soon after the outburst peak, the profiles obtained with \insight{} showed a two-peak pattern, implying a fan-beam emission geometry, when the X-ray luminosity is so high that the radiation pressure could stop the accreting matter above the surface via radiation-dominated shock \citep{chen2008study}. While at the low luminosity level, the profiles detected by \insight{} suggest a pencil-beam geometry with profiles having an one-peak pattern. \cite{becker2012spectral} argued that the critical luminosity at which the emitting pattern changes depends on the NS magnetic field as 
\begin{equation}
    L_{\mathrm{crit}} = 1.5 \times 10^{37} B_{12}^{16/15}\,\mathrm{erg}\,\mathrm{s}^{-1}.
    \label{eq:lcrit}
\end{equation} 
The observed transition of the pulse profile shape happens around $4\times10^{-9}\,\mathrm{erg}\,\mathrm{cm}^{2}\,\mathrm{s}^{-1}$ according the \insight{} data. Using this value one can plot possible distances and magnetic field values satisfying Equation \ref{eq:lcrit} (see solid line in Figure \ref{fig:distance}). The blue region corresponds to the uncertainty of the critical flux. The red circle in Figure \ref{fig:distance} suggests the distance and the magnetic field based on both the torque model and critical luminosity. The resulting distance is $12.4\pm0.2\,\mathrm{kpc}$ and the magnetic field is $\sim 4.3^{+0.6}_{-0.5}\times 10^{12}\,\mathrm{G}$.

\par 

We note that the magnetic field values inferred from the possible cyclotron lines either at $\sim10$\,keV or $\sim30$\,keV are of the same order as derived above. 
\cite{shi2015super,wang1987disc} argued that the magnetic field is overestimated by the \cite{ghosh1979accretion}, and the $L_\mathrm{crit}$ is highly uncertain \citep{becker2012spectral, mushtukov2015critical}. 

Nevertheless, the torque model and the variation of the pulse profile provide a new measure of the distance ($12.4\pm0.2\,\mathrm{kpc}$), which is consistent with the optical observation \citep{strader2019optical}. 

Also further observation of the propeller effect will provide a new constrain to the distance and the magnetic field in 4U 1901+03. 

\section{Acknowledgement}
This work made use of the data from the \insight{} mission, a project funded by the China National Space Administration (CNSA) and the Chinese Academy of Sciences (CAS). We gratefully acknowledge the support from the National Program on Key Research and Development Project (grant No. 2016YFA0400801) from the Minister of Science and Technology of China (MOST) and the Strategic Priority Research Program of the Chinese Academy of Sciences (grant No. XDB23040400). The authors are thankful for support from the National Natural Science Foundation of China under grande Nos. 11673023, 11733009, U1838108, U1838201, U1838202, and U1938103; and Russian Science Foundation grant 19-12-00423 (SST). 

\section*{References}

\bibliography{reference.bib}

\end{document}